\documentclass[article]{IEEEtran} \newdimen\figwidth  \figwidth=9cm

\usepackage[english]{babel}
\usepackage{epsfig}
\usepackage{amsmath}
\usepackage{cite}
\usepackage{balance}
\usepackage{amssymb}
\usepackage[center]{caption}

\newtheorem{theorem}{Theorem}
\newtheorem{corollary}{Corollary}
\newtheorem{lemma}{Lemma}
\newcommand{\avr}{\textbf{E}}

\begin{document}

\title{On the Capacity of Fading Channels with Peak and Average Power Constraints at Low SNR}

\author{
\IEEEauthorblockN{\large Longguang Li$^{\text{1}}$, Lokman Sboui$^{\text{2}}$, Zouheir Rezki$^{\text{3}}$,  and Mohamed-Slim Alouini$^{\text{4}}$}\\[.2cm]
\IEEEauthorblockA{
\small $^{\text{1}}$Telecom ParisTech, Paris, France.\\
\small $^{\text{2}}$\'Ecole de Technologie Sup\'erieure (\'ETS), Montreal, Canada \\
$^{\text{3}}$University of Idaho, Moscow, ID, USA.\\
$^{\text{4}}$King Abdullah University of Science and Technology (KAUST),Thuwal, Makkah Province, Saudi Arabia.\\
\small  longguang.li@telecom-paristech.fr, lokman.sboui@lacime.etsmtl.ca, zrezki@uidaho.edu, slim.alouini@kaust.edu.sa}}

\maketitle

\begin{abstract}
The capacity of fading channels under peak and average power constraints in the low-SNR regime is investigated. We show that the capacity scales essentially as   ${C \approx A \ \text{SNR}  \int_{1- \frac{1}{A}}^1 F^{-1}\left(t\right)dt}$, where $A$ is the peak to average power ratio (PAPR), and  $F(\cdot)$ is the cumulative distribution function of the fading channel.
 We also prove that an On-Off power scheme is sufficient to asymptotically achieve the capacity. Furthermore, by considering the variable PAPR scenario, we generalize the scalability of the capacity and derive the asymptotic expression for the capacity in the low-SNR regime.
 \end{abstract}

\section{Introduction}\label{S1}
In response to the increasing demand for higher power efficiency in wireless communications, many researchers have been devoted to setting up the theory on performance limits in the power limited systems~\cite{Verdu2002,Sboui2013b, Sboui2014b}. Correspondingly, many new practical power allocation schemes have been proposed to approach these limits. For instance, different adaptive schemes
  take advantage of the feedback channel to increase the power efficiency,~\cite{Yao2005,Sboui2017}.
 The low-SNR framework does not apply only to applications where the power budget is asymptotically low, but also includes applications where the power budget is arbitrary and the available degrees of freedom (DoF) are large enough. For instance, many wideband wireless communication systems, e.g., satellite, deep-space, and sensor networks can achieve a very high total rate by utilizing large DoF.

The capacity of fading channels in the low-SNR regime has been extensively studied in the current literature. One aspect of this study is on the analysis of the low-SNR capacity limit \cite{shannon48_1}. In \cite{kennedy_69}, it first shows that the capacity of a Rayleigh-fading channel shares the same limit with that of an additive white Gaussian noise (AWGN). This result is famously generalized in \cite{telatar_00} by showing that this limit can be achieved with arbitrary fading channels.  The works \cite{medard_02,verdu_02} subsequently show that some bursty signaling suffices to achieve the capacity limit of fading channels. Another aspect is on the scaling of capacity with SNR variations. In~\cite{Borade2010} it has been shown that the capacity of Rayleigh fading channels with perfect channel state information (CSI) at the transmitter side (CSI-T) and at the receiver side (CSI-R) scales essentially as $\text{SNR}\, \log{\frac{1}{\text{SNR}}}$. The works in~\cite{Rezki2012} generalize the channel model to Nakagami-m channels. The capacity is shown to behave as $\frac{\Omega}{m} \, \text{SNR} \, \log{\frac{1}{\text{SNR}}}$, where $m$ is the Nakagami-m fading parameter and $\Omega$ is the channel mean-square. More advances in the field can be found in the survey \cite{porrat_07}.

The main limitation of these previous characterizations at low SNR lies in their high peak to average power ratio (PAPR). In fact, as SNR goes to zero, the PAPR approaches infinity. In practical communication systems, the PAPR is limited by hardware restrictions. In addition, as the PAPR is larger, the RF power amplifier is required to operate in the high back-off region resulting in increased cost. Consequently, it is important to consider these practical constraints~\cite{Tarokh2000}. However, imposing the peak power constraint into the channel input makes the capacity characterization in certain channels become extremely difficult. A first seminal work by Smith in \cite{smith_71} surprisingly shows that the capacity-achieving input distribution in the AWGN channel turns out to be a probability mass function with finite support. The results were subsequently extended to the complex Gaussian channel \cite{shamai_95}, Rayleigh fading channel \cite{abou_01}, and Rician fading channel \cite{gursoy05_1,gursoy05_2}. However, exactly characterizing the capacity of the peak power imposed Gaussian channel is still insurmountable. Instead, the works \cite{dytso_17,lapidoth_09, moser_18, chaaban18_1, chaaban18_2} derive asymptotically tight lower and upper bounds when there are peak or both peak and average power constraints in the Gaussian (or Gaussian-like) channel. In this paper, we are interested in the capacity analysis of fading channels in the low-SNR regime when there is both peak and average power constraints.  The related works in \cite{ subramanian_02} characterize the asymptotic capacity with neither CSI-T nor CSI-R.

Our main goal, in this letter, is to better understand the capacity of communication under peak and average power constraints in the low-SNR regime. We investigate the capacity of fading channels subject to both peak and average power constraints with perfect CSI at both the transmitter and the receiver (CSI-TR).

Our contributions can be summarized as follows:
\begin{itemize}
\item Analyzing the effect of an additional peak power constraint $A$ on the optimal power profile.
\item Proving that under both peak and average power constraints, the capacity scales essentially as   $ {C \approx A \ \text{SNR} \underset{\left|{h}\right|^{2} \ge \lambda }{\avr} {\left[ \left|{h}\right|^{2}\right]} }$, where $\lambda$ is the Lagrange multiplier associated to the average power constraint.
\item Presenting an On-Off power scheme which is asymptotically optimal. This implies that in the low-SNR regime, a 1-bit feedback on the CSI-T is enough to approach the channel capacity, which is very helpful from a signaling design point of view.
\item Studying the case where the PAPR constraint depends on SNR, and generalize the scalability of capacity in the low-SNR regime. Specializing our result to Rayleigh fading, we derive a simple asymptotic expression of the capacity.
\end{itemize}

\section{System Model}\label{S2}
We consider the continuous complex-baseband model of a flat fading single-input single-output (SISO) communication specified by
\begin{equation}\label{E201}
  {y} = {h} \, {x}+{v},
\end{equation}
where ${y}$, ${h}$, and ${x}$  are complex random variables (r.v.) representing the received signal, the channel gain, and the transmitted signal, respectively.  The r.v. ${v}$ represents a circularly symmetric complex Gaussian noise  with mean zero, and  unit variance, i.e., ${v} \sim \mathcal{CN}\left(0, 1\right)$ and is independent of all other r.v.
We assume perfect CSI-TR and that the channel follows an arbitrary continuous fading distributions. The average transmitted power is constrained to
$\underset{p_{{x}|h} }{\avr}\left[P\left(h\right)\right] \leq \text{SNR}$,
and the peak power at the transmitter side is restricted to
\begin{equation}
\underset{h}{{\max}} \ P\left(h\right)\leq A \ \text{SNR},
\end{equation}
 where $A\geq1$ is the PAPR, and the previous averaging is over all input conditioned distributions $p_{{x}}|h$. In the following sections, the notation ${f\left(x\right) \approx g\left(x\right)}$ will be used to denote
 \begin{equation}
 \underset{\text{SNR} \rightarrow 0}\lim{\frac{f\left(\text{SNR}\right)}{g \left(\text{SNR}\right)}} = 1.
 \end{equation}
 Inequalities $\tilde{<}$ and $\tilde{>}$ are defined similarly.

\section{Low SNR Capacity With Perfect CSI-TR}\label{S3}
In this section, we briefly recall results  on the channel capacity.  Then, we present our main result, namely, the asymptotic behavior of capacity subject to both peak and average power constraints, along with the corresponding proof. Then, we show that when SNR goes to zero, the optimal power allocation can be achieved (asymptotically) by an On-Off power scheme.
\subsection{Capacity Results with no PAPR Constraint}
With perfect CSI-TR, the ergodic capacity subject to an average power constraint is maximized by the well-known water-filling scheme, $P\left(h\right)$, given by:
\begin{equation}\label{EQ02}
P\left(h\right)=\left[\frac{1}{\lambda_0}-\frac{1}{\left|h\right|^{2}}\right]^{+},
\end{equation}
where $\left[x\right]^{+}=\max{\left\{0, x\right\}}$, and where $\lambda_0$ is the Lagrange multiplier obtained by satisfying the average power constraint with equality \cite{Goldsmith1997}, i.e.,
\begin{equation}
\text{SNR}=\underset{\left|{h}\right|^{2}}{\avr}\left[\left[\frac{1}{\lambda_0}-\frac{1}{\left|h\right|^{2}}\right]^{+}\right].
\end{equation}

The capacity is then obtained by averaging $\log{\left(1+\text{SNR}\left| {h}\right|^{2}\right)}$, and is given by:
\begin{equation}\label{E301}
C=\int_{\lambda_0}^{\infty}{\log{\left(\frac{t}{\lambda_0}\right)f_{\left|{h}\right|^{2}}(t) \, dt}}.
\end{equation}

Note that this result does not account for the peak power constraint. Since the PAPR needs to be taken into consideration, we analyze the PAPR  in this classic optimal power allocation. From the expression of $P\left(h\right)$ in~(\ref{EQ02}), the PAPR, denoted as  $\mathcal{A}\left(\text{SNR}\right)$, can be~obtained~as
\begin{equation}
\mathcal{A}\left(\text{SNR}\right)=\frac{1}{\lambda_0 \cdot  \text{SNR}}.
\end{equation}

Thus, $\mathcal{A}\left(\text{SNR}\right)$ only depends on SNR. To analyze the properties of PAPR, we define the function $\mathcal{A}\left(x\right)$ on $(0, \infty)$~as:
\begin{equation} \label{E302}
 \mathcal{A}\left(x\right) = \frac{1}{x\underset{\left|h \right |^2} {\avr} \left[  \left[ \frac{1}{x} - \frac{1}{\left|h\right|^2} \right]^{+}   \right] },
\end{equation}
and we summarize its properties  in Lemma~\ref{L1}.

\begin{lemma}\label{L1}
The function $\mathcal{A}\left(x\right)$ defined by (\ref{E302}) is
\begin{enumerate}
\item continuous and strictly monotonically increasing;
\item when $x \rightarrow 0$, $\mathcal{A}\left(x\right) \rightarrow 1$;
\item when $x \rightarrow \infty$, $\mathcal{A}\left(x\right)\rightarrow \infty$.
\end{enumerate}
\end{lemma}

\begin{IEEEproof}
The continuous property in part 1) can be directly derived by construction from (\ref{E302}). Since its first derivative
\begin{equation}
{\mathcal{A}^{'}\left(x\right) = -\frac{-\underset{\left|h\right|^2}{\avr}\left[ \frac{1}{\left|h\right|^2}\right]}{\left\lbrace x\underset{\left|h \right |^2} {\avr} \left[  \left[ \frac{1}{x} - \frac{1}{\left|h\right|^2} \right]^{+}   \right]  \right\rbrace^{2} } < 0},
\end{equation}
  $\mathcal{A}\left(x\right)$ is monotonically increasing.

Part 2) can be immediately established by seeing that when $x$ goes to zero, we have:
\begin{equation}
\mathcal{A}\left(x\right) \approx \frac{1}{\underset{\left|h\right|^2}{\avr}{\left[ \left[ 1-\frac{x}{\left|h\right|^2}  \right]^{+} \right]}} \approx 1.
\end{equation}

To prove part 3), since $\mathcal{A}\left(x\right) > \frac{1}{1-F_{\left|h\right|^2}\left(x\right)}$, and when $x \rightarrow \infty$, $\frac{1}{1-F_{\left|h\right|^2}\left(x\right)} \rightarrow \infty$. Therefore, $\mathcal{A}\left(x\right) \rightarrow \infty$ when $x \rightarrow \infty$.
\end{IEEEproof}

Lemma~\ref{L1} shows that the PAPR approaches 1 in high SNR, which implies that PAPR is not crucial in this regime and there is little benefit in adapting the power in high SNR in agreement with the results in \cite{Goldsmith1997}. However, in the low-SNR regime, in order to achieve the capacity, as SNR goes to zero, the PAPR goes to infinity, which is difficult to implement in practical communication systems. Therefore,  considering the effect of the PAPR constraint on the capacity in the low-SNR regime is important when designing low-SNR systems.

\subsection{Constant PAPR Constraint}
\subsubsection{Capacity Results}
In~\cite{Khojastepour2004}, the optimal power allocation subject to both peak and average power constraint. The optimal power, $P\left(h\right)$ is given~by:
\begin{equation}
 P\left(h\right) = \min\left\{ \left[ \frac{1}{\lambda} -\frac{1}{\left|h\right|^2} \right]^{+}, \ A \ \text{SNR} \right\}
\end{equation}
where $\lambda$ is the Lagrange multiplier obtained by satisfying the average power constraint with equality:
\begin{equation}\label{E303}
\text{SNR}=\underset{\left|{h}\right|^{2}}{\avr}\left[\min\left\{\left[\frac{1}{\lambda}-\frac{1}{\left|h\right|^{2}}\right]^{+}, \ A \ \text{SNR}\right\} \right].
\end{equation}
The corresponding capacity is given by
\begin{equation} \label{E304}
C\left(\lambda\right)=\int_{\lambda}^{\infty}{ \min\left\{ \log{\frac{t}{\lambda}}, \log{\left( 1+A \ \text{SNR} \ t \right)} \right\} f_{\left|{h}\right|^{2}}\left(t\right)dt.   }
\end{equation}

In the following Lemma~\ref{L2}, we first present  some properties of $\lambda$, which will be useful in deriving the asymptotic capacity.
\begin{lemma}\label{L2}
The Lagrange multiplier  $\lambda$ satisfies the following proprieties:
\begin{enumerate}
\item $\lambda$ is restricted into the following interval:
\begin{equation}
\lambda \in \left[ \frac{F^{-1}\left(1-\frac{1}{A}\right)}{1+A \ \text{SNR}F^{-1}\left( 1-\frac{1}{A}\right)}, F^{-1}\left( 1- \frac{1}{A}\right)  \right],
\end{equation}
where $F^{-1}\left(\cdot\right)$ represents the inverse function of the cumulative distribution function of $f\left(\cdot\right)$;
\item when $\text{SNR} \rightarrow 0, \lambda \rightarrow F^{-1}\left(1-\frac{1}{A}\right)$;
\item when $\text{SNR} \rightarrow 0, \lambda^{n} \text{SNR} \rightarrow 0$, where $n$ is an arbitrary positive constant.
\end{enumerate}
\end{lemma}

\begin{IEEEproof}
For part 1), we first note that $\frac{1}{\lambda}- A \ \text{SNR}  > 0$ by the fact that $\frac{1}{\lambda}$ should be lower than the power  peak given by $A \ \text{SNR}$.
Then, we explicitly express the optimal power, $P\left(h\right)$,~as:
\begin{equation}\label{E305}
P\left(h\right) =
  \begin{cases}
   0, & \text{if} \left|{h}\right|^{2} \leq \lambda \\
   \frac{1}{\lambda}-\frac{1}{\left|{h}\right|^{2}},  & \text{if }  \lambda < \left|{h}\right|^{2} < \left[\frac{\lambda}{1-\lambda \ A \ \text{SNR}}\right]^{+} \\[3mm]
   A \ \text{SNR}, & \text{if}  \left|{h}\right|^{2} \geq   \frac{\lambda}{1-\lambda \ A \ \text{SNR}}.
  \end{cases}
\end{equation}
Note that when $\lambda < \left|{h}\right|^{2} <   \frac{\lambda}{1-\lambda \ A \ \text{SNR}}  $, we have
\begin{equation}
0 \leq \frac{1}{\lambda}-\frac{1}{\left|{h}\right|^{2}} \leq A \ \text{SNR}.
\end{equation}
Using $0$ and $A \ \text{SNR}$ to substitute $\frac{1}{\lambda}-\frac{1}{\left|{h}\right|^{2}}$, respectively, the following inequalities hold:
\begin{align}
\underset{\left|{h}\right|^{2} > \frac{\lambda}{1-\lambda \ A \ \text{SNR}}}{\avr}& {\left[A \ \text{SNR} \right]} \nonumber\\
&\hspace{-1.5cm}\leq \underset{\left|{h}\right|^{2}}{\avr}\left[\min\left\{\left[\frac{1}{\lambda}-\frac{1}{\left|h\right|^{2}}\right]^{+}, A \ \text{SNR}\right\} \right] \\
&\hspace{-1.5cm}\leq \underset{\left|{h}\right|^{2} > \lambda}{\avr} {\left[A \ \text{SNR} \right]}.
\end{align}
The above inequalities can be equivalently expressed as
\begin{align}
1-F\left( \frac{\lambda}{1-\lambda \ A \ \text{SNR}} \right) \leq \frac{1}{A} \leq   1- F\left( \lambda \right).
\end{align}
By further simplification, we obtain:
\begin{align}
\frac{F^{-1}\left(1-\frac{1}{A}\right)}{1+A \ \text{SNR}F^{-1}\left( 1-\frac{1}{A}\right)}  \leq \lambda \leq F^{-1}\left(1- \frac{1}{A}\right).
\end{align}

 Part 2) can be established by seeing that as $\text{SNR} \rightarrow 0$, $\frac{F^{-1}\left(1-\frac{1}{A}\right)}{1+A \ \text{SNR}F^{-1}\left( 1-\frac{1}{A}\right)} \rightarrow F^{-1}\left(1- \frac{1}{A}\right)$, then the lower bound coincides with the upper bound.

Part 3) is straightforward by seeing that $\lambda$ lies in a bounded interval.
\end{IEEEproof}
Note that $1-\frac{1}{A}\geq0$ since peak power is greater or equal to the average power.
As we can see in Lemma~\ref{L2}, when $A$ is fixed, $\lambda$ is upper bounded by $F^{-1}\left(1- \frac{1}{A}\right)$, which does not scale with SNR. Next, we state our main theorem.
\begin{theorem}\label{T1}
For fading channels with perfect {CSI-TR}, the asymptotically low SNR~capacity~is
\begin{equation}
C \approx  A \ \text{SNR}  \int_{1- \frac{1}{A}}^1 F^{-1}\left(t\right)dt.
\end{equation}
\end{theorem}
\begin{IEEEproof}
The proof of Theorem~\ref{T1} is technical, so we defer it to Appendix A.
\end{IEEEproof}

As shown in Theorem 1, when the PAPR constraint is adopted, the capacity at the low SNR regime linearly changes with SNR. The PAPR constraint has much effect on the achievable capacity. In the special case of $A = 1$, the power scheme evolves to be the constant power allocation, $C \approx \text{SNR}$. When~$A$ approaches infinity, the power scheme becomes the classic water-filling power allocation. In the scenarios of limited system power, increasing the PAPR is an effective way to improve the achievable capacity.

\subsubsection{On-Off  Power Scheme}
We note that an On-Off power scheme archives the capacity as described in the following Corollary.
\begin{corollary}\label{C1}
An On-Off power scheme that transmits when $\left|h\right|^{2} \geq  F^{-1}\left(1- \frac{1}{A}\right)$, with a power $A \ \text{SNR}$, and remains silent otherwise, is asymptotically optimal.
\end{corollary}

\begin{IEEEproof}
The On-Off power scheme can be explicitly expressed as:
\begin{equation}\label{E332}
P\left(h\right)=
\begin{cases}
 A \ \text{SNR}, & \text{if} \, \left|h\right|^{2} \geq  F^{-1}\left( 1 - \frac{1}{A} \right) \\
    \ \ \ 0, & \text{otherwise}.
\end{cases}
\end{equation}

It is clear that $P\left(h\right)$ in \eqref{E332} is an eligible candidate since it satisfies the average and peak power constraint. Following is the proof of the rate achieved by the scheme, which asymptotically approaches the capacity.

According to Lemma~$2$, $\lambda \geq   \frac{F^{-1}\left(1-\frac{1}{A}\right)}{1+A \ \text{SNR}F^{-1}\left( 1-\frac{1}{A}\right)} $ which yields to $ F^{-1}\left(1-\frac{1}{A}\right) \leq \frac{\lambda}{1- \lambda \ A \ \text{SNR}}$.

The rate achieved by the On-Off power scheme satisfies:
\begin{align}
R &= \int_{F^{-1}\left(1-\frac{1}{A}\right)}^{\infty}{\log\left(1 + A \ \text{SNR} \ t \right) f_{\left|{h}\right|^{2}}\left(t\right) dt} \\
  &\geq \int_{\frac{\lambda}{1-\lambda \ A \ \text{SNR}}}^{\infty}{\log\left(1 + A \ \text{SNR} \ t \right) f_{\left|{h}\right|^{2}}\left(t\right) dt} \\
  & \approx \int_{\lambda}^{\infty}{\log{\left(1+A \ \text{SNR} \ t\right)}f_{\left|{h}\right|^{2}}\left(t\right)dt }\\
  &\approx A \ \text{SNR}   \int_{1- \frac{1}{A}}^\infty  F^{-1}\left(t\right)dt.
\end{align}
The proof is concluded.
\end{IEEEproof}
The asymptotic optimality of the On-Off power in Corollary~\ref{C1} implies that only one bit of CSI-T at each fading realization is required. This bit contains the result of comparing $\left|h\right|^{2}$ and $F^{-1}\left(1-\frac{1}{A}\right)$. We also present a different way proving the optimality of the On-Off power scheme as follows.
The average power constraint (\ref{E303}) can be explicitly expressed as:
\begin{equation}
\text{SNR}=\hspace{-2mm}\int_{\lambda}^{\frac{\lambda}{1-\lambda \ A \ \text{SNR}}}\hspace{-2mm}{\left(\frac{1}{\lambda}-\frac{1}{t} \right)f\left(t\right) dt} + \int_{\frac{\lambda}{1-\lambda \ A \ \text{SNR}}}^{\infty}\hspace{-6mm}{A \ \text{SNR}  f\left(t\right) dt}.
\end{equation}
The first term corresponds to the power allocation requiring the exact CSI-T. As SNR decreases, it scales~as
\begin{align}
&\int_{\lambda}^{\frac{\lambda}{1-\lambda \ A \ \text{SNR}}}  {\left(\frac{1}{\lambda}-\frac{1}{t} \right)f\left(t\right) dt} \, \, \nonumber \\
&\tilde{<}  \, \,  \underset{t \in \left[ \lambda, \frac{\lambda}{1-\lambda \ A \ \text{SNR}} \right]} {\max} f\left( t\right)  A \ \text{SNR} \left(\frac{\lambda}{1-\lambda \ A \ \text{SNR}} -\lambda \right) \\
&\approx f\left(\lambda \right)  \lambda^{2} A^2 \text{SNR}^2 \\
&\approx o \left( \text{SNR}\right).
\end{align}
The second terms belongs to the constant power allocation, which behaves as
\begin{align}
  \int_{\frac{\lambda}{1-\lambda \ A \ \text{SNR}}}^{\infty}{A \ \text{SNR}f\left(t\right) dt}
 &\approx \int_{\lambda}^{\infty}{A \ \text{SNR}f\left(t\right) dt}  \\
 &\approx A \ \text{SNR}  \left[ 1-F\left(\lambda\right)  \right].
\end{align}

When $\text{SNR} \rightarrow 0$, the second part dominates. Therefore, one bit of CSI-T is sufficient.
We apply our results to Rayleigh fading channels, and we obtain the results in Corollary~\ref{C2}.
\begin{corollary}\label{C2}
For the Rayleigh fading channel described in~\cite{Borade2010}, the capacity is given by:
\begin{equation} \label{E306}
C \approx \text{SNR} \left( 1+ \log{A} \right).
\end{equation}
\end{corollary}
\begin{IEEEproof}
For Rayleigh fading, $ F^{-1}\left(1-\frac{1}{A}\right) = \log{A}$. Since the rate achieved by the On-Off power scheme proposed in Corollary~\ref{C1} is asymptotically optimal, the capacity can be derived as:
\begin{align}
C \approx R  &= \int_{\log{A}}^{\infty}{\log\left(1 + A \ \text{SNR} \ t \right) e^{-t} dt} \\
  &\approx A \ \text{SNR} \int_{\log{A}}^{\infty}{t e^{-t} dt} \\
  &= \text{SNR} \left(1 + \log{A}\right).
\end{align}
The proof is concluded.
\end{IEEEproof}
\subsubsection{Energy Efficiency }
We also characterize the energy efficiency of arbitrary fading channels at low SNR.
To illustrate this observation, we denote by  $E_{n}$  the transmitted energy in Joules per information nats. Then, we have: $\frac{E_{n}}{\sigma_{v}^{2}} \; C^{0}\left(\text{SNR}\right) = \text{SNR}$. By using Theorem~1, we obtain:
\begin{equation}\label{E331}
\frac{E_{n}}{\sigma_{v}^{2}} \approx \frac{1}{A \int_{1- \frac{1}{A}}^\infty F^{-1}\left(t\right)dt },
\end{equation}
which indicates that the energy required to communicate one nat of information reliably only depends on the PAPR $A$. One can achieve a lower energy efficiency per information bit by increasing $A$. In the special case of Rayleigh fading, we have $\frac{E_{n}}{\sigma_{v}^{2}} \approx \frac{1}{1+\log{A} }$.

\subsection{Variable PAPR Constraint}
This section considers the case where the PAPR varies with the SNR. In practical communications systems, the PAPR may not be fixed as SNR decreases. Also, exploring this area offers a general formula on the optimal power allocation under both peak and average power constraints at low SNR. To simplify the analysis, we focus on Rayleigh fading channels in this section.
The variable PAPR is denoted as $\mathcal{A}\left(\text{SNR}\right)$. Some properties, that the effective $\mathcal{A}\left(\text{SNR}\right)$ needs to satisfy, are summarized in the Lemma~\ref{L3}.
\begin{lemma}\label{L3}
The effective $\mathcal{A}\left(\text{SNR}\right)$ should satisfy:
\begin{enumerate}
\item When $\text{SNR} \rightarrow 0$, $\mathcal{A}\left(\text{SNR}\right) \text{SNR} \rightarrow 0$;
\item When $\text{SNR} \rightarrow 0$, if $\mathcal{A}\left(\text{SNR}\right) \rightarrow \infty$, then $\lambda \rightarrow \infty$.
\end{enumerate}
\end{lemma}

\begin{IEEEproof}
To prove 1), from section A, we know that the peak power without PAPR constraint is $\frac{1}{\lambda_0}$. Therefore, to make the PAPR constraint effective, a necessary condition is  $\mathcal{A}\left(\text{SNR}\right) \text{SNR} < \frac{1}{\lambda_0}$. Since $\frac{1}{\lambda_0} \rightarrow 0$, when $\text{SNR} \rightarrow 0 $,  then $\mathcal{A}\left(\text{SNR}\right) \text{SNR}$ must approach $0$.

To prove part 2), from the average power constraint, we have
\begin{equation}
 \int_{\frac{\lambda}{1-\lambda \ \mathcal{A}\left(\text{SNR}\right) \text{SNR}}}^{\infty}{\mathcal{A}\left(\text{SNR}\right) \text{SNR}f_{\left|{h}\right|^{2}}(t) \, dt} \leq  \text{SNR},
\end{equation}
which, after simplifications, gives
\begin{equation}
\frac{1}{\lambda} < \mathcal{A}\left(\text{SNR}\right) \text{SNR} +\frac{1}{F^{-1}\left(1-\frac{1}{A}\right)}.
\end{equation}
Since the RHS of the equality goes to $0$,we have $\lambda \rightarrow \infty$, as $\mathcal{A}\left(\text{SNR}\right) \rightarrow \infty$.
\end{IEEEproof}

To analyze the effect of the variable PAPR constraint on the capacity, we assume that $A(\text{SNR})$ satisfies all the conditions in Lemma~\ref{L3}. Then, we present the main theorem of the capacity with variable PAPR constraint in the low SNR regime.

\begin{theorem}\label{T2}
Let
\begin{equation}
l\left(\text{SNR}\right)=\frac{\lambda \ \mathcal{A}\left(\text{SNR}\right) \text{SNR} }{\frac{1}{\lambda}- \mathcal{A}\left(\text{SNR}\right) \text{SNR} }  = \frac{1}{\frac{1}{\lambda} -  \mathcal{A}\left(\text{SNR}\right)\text{SNR} } - \lambda,
\end{equation}
and define $l_{0}$ as
\begin{equation}
 l_{0} = \underset{\text{SNR} \rightarrow 0}\lim {l}.
 \end{equation}
 Then, for Rayleigh fading channels with variable PAPR constraint, the capacity can be asymptotically expressed as
\end{theorem}
\begin{equation}\label{E400}
C \approx
\begin{cases}
\text{SNR} \log\left({\frac{1}{\text{SNR}} }\right), &\text{if} \, \, l_{0}  >   0\\[3mm]
 \text{SNR} \log\left({ \mathcal{A}\left(\text{SNR}\right) }\right), & \text{if} \, \, l_{0} = 0.
\end{cases}
\end{equation}
\begin{IEEEproof}
See Appendix B.
\end{IEEEproof}

Intuitively, the value of $l$ corresponds to the region where the transmit power is adapted with the feedback of channel gain. The other regions belong to the silent mode or constant transmitting power mode. When $l$ goes to infinity, the power allocated to this region dominates. On the other hand, when $l$ approaches $0$, the power allocated to the constant power mode dominates.

The expressions in \eqref{E400} seem very different in two cases where $l_{0}  >   0$ and $l_{0}  >   0$. In fact, it's not. In the following, we show that when $l_{0} >0$,
\begin{equation}
C \approx \text{SNR} \log\left({\frac{1}{\text{SNR}} }\right) \approx \text{SNR} \log\left({ \mathcal{A}\left(\text{SNR}\right) }\right).
\end{equation}

To see this, from the definition of $l$, we first obtain the following equation:
\begin{equation}\label{E340}
\mathcal{A}\left(\text{SNR}\right) = \frac{1}{\text{SNR}} \  \frac{1}{\lambda} \  \frac{l}{l+\lambda}.
\end{equation}
From the proof in Appendix B, we know that $\lambda \approx \log {\frac{1}{\text{SNR}}  }$. substituting $\lambda$ into \eqref{E340}, we have

\begin{equation}\label{E341}
\frac{1}{\text{SNR}}  {\left(  \frac{1}{\log{\frac{1} {\text{SNR}}}} \right)  }^{2} \, \, \tilde{<}  \, \,  \mathcal{A}\left(\text{SNR}\right) \, \, \tilde{<}  \, \, \frac{1}{\text{SNR}}  \frac{1}{\log{\frac{1} {\text{SNR}}}}.
\end{equation}
It's immediate to see
\begin{equation}\log{\mathcal{A}\left(\text{SNR}\right)} \approx \log{\frac{1}{\text{SNR}}}.\end{equation}
Therefore, in both cases, the capacity can be expressed as
\begin{equation}
C \approx \text{SNR} \ \mathcal{A}\left(\text{SNR}\right).
\label{eq:capacity}
\end{equation}
Moreover, \eqref{eq:capacity} even holds for the channel without peak power constraint in the low-SNR regime. As Lemma~1 in~\cite{Rezki2012} shows that the PAPR in channels without peak power constraint is
\begin{align}
\mathcal{A}\left(\text{SNR}\right)  = \frac{1}{\lambda \ \text{SNR}} \approx \frac{1}{\text{SNR}  \log{\frac{1}{\text{SNR}} }}.
\end{align}
The capacity in~\cite{Rezki2012} is then given by:
\begin{align}
C &\approx \text{SNR} \log{\frac{1}{\text{SNR}}}  \\
 &\approx \text{SNR}  \log{\frac{1}{\text{SNR}\log{\frac{1}{\text{SNR}}}}}\\
 &\approx \text{SNR} \log{\mathcal{A}\left(\text{SNR}\right)}.
\end{align}
This shows that \eqref{eq:capacity} always hold in the low-SNR regime, irrespective of the activeness of the imposed peak power constraint. Hence, we get a new perspective on the capacity of Rayleigh fading channels in the low-SNR regime. The asymptotic capacity are actually only determined by two parameters: SNR and $\mathcal{A}\left(\text{SNR}\right)$.

\section{Numerical Results and Discussion}\label{S5}
\begin{figure}
\begin{center}
\includegraphics[width=\figwidth]{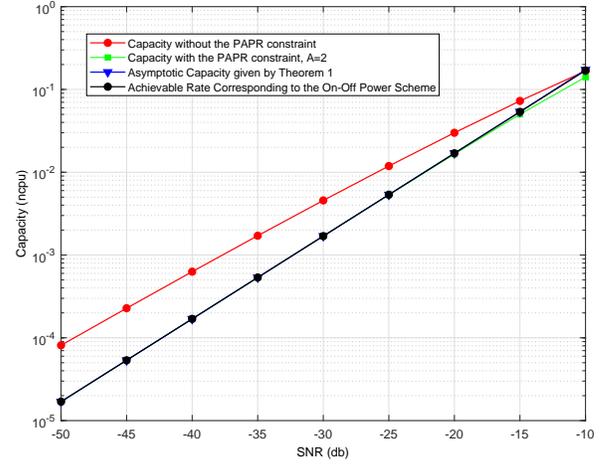}
\end{center}
\caption{Capacity versus SNR with Rayleigh fading with $A=2$.}
    \label{F1-2}
\end{figure}
We first consider the constant PAPR in Rayleigh fading channel. Figure~\ref{F1-2} depicts the ergodic capacity in nats per channel use (npcu). The optimal power allocation with and without the PAPR obtained by using standard optimization methods are presented as the benchmark curves. In Fig.~\ref{F1-2},  where  PAPR=$2$, we show that the gap between the two curves increases as SNR goes to $0$  implying that the PAPR constraint has a higher impact on the capacity at low SNR. Also, the asymptotic capacity representing the low SNR characterization given by Theorem~1 is shown in the figure. The asymptotical capacity curve accords well with the capacity curve with the PAPR constraint.
Furthermore, the capacity of the proposed On-Off scheme is plotted in Fig.~\ref{F1-2}. This rate matches perfectly the exact curve at the low SNR values displayed in Fig.~1.

\begin{figure}
\begin{center}
\includegraphics[width=\figwidth]{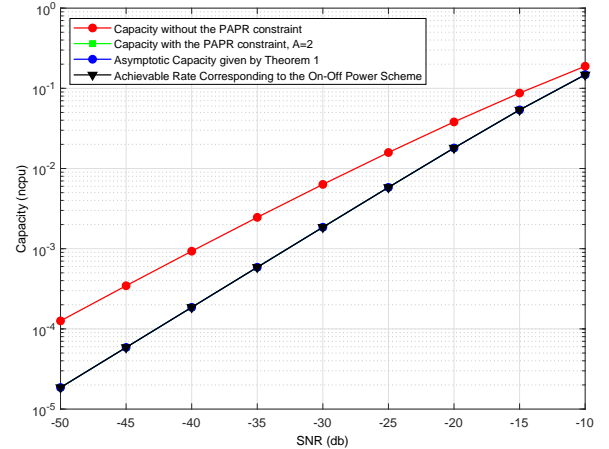}
\end{center}
\caption{Capacity versus SNR with Nakagami-m fading with $m=\frac{1}{2}$.}
    \label{F3}
\end{figure}
Figure~\ref{F3}  displays the capacity of an i.i.d. Nakagami-m fading channel for $m=\frac{1}{2}$.
As can be seen in both figures, the curves depicting the characterizations in Theorem~\ref{T1} follow the same shape as the curve obtained by optimal power allocation with the PAPR constraint. In addition, the On-Off scheme achievable rate is also depicted and is almost indistinguishable from the capacity curve showing that the proposed suboptimal scheme is accurate at the low SNR regime.
\begin{figure}[h]
\begin{center}
\includegraphics[width=\figwidth]{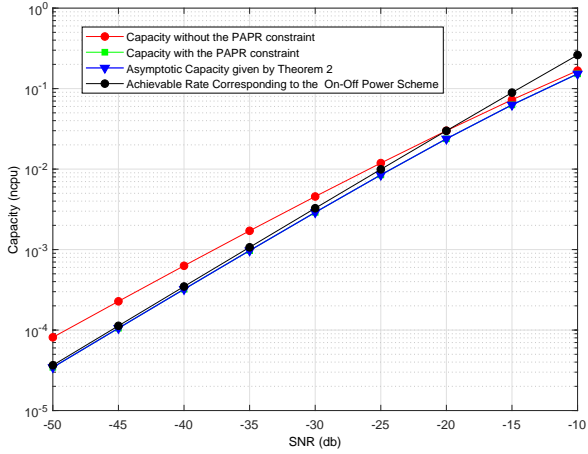}
\end{center}
\caption{Capacity versus SNR   for $l_{0}=0$ and $l_{0}>0$ with  $l_{0}=0$.}
    \label{F4}
\end{figure}

In Fig.~\ref{F4}, we consider the variable PAPR scenario following  Lemma~\ref{L3}. In Fig.~\ref{F4}, we choose PAPR=$\log{\left(e+\frac{1}{\text{SNR}}\right)}$, which satisfies the requirement on trend of the PAPR. Also, the PAPR corresponds to  $l_0 = 0$ scenario in Eq.~(\ref{E400}). We note that the curve obtained by Eq.~(\ref{E400}) matches very well  the capacity obtained by the optimal power allocation. Furthermore, the On-Off power schemes achieve good capacity approximation for SNR below $-35$ dB.

\section{Conclusion}
In this paper, we studied the capacity of fading channels at low SNR under peak and average power constraints. We presented an approximation of the capacity as a function of the SNR and the peak to average power ratio (PAPR). We also presented a practical On-Off scheme that achieves asymptotical optimal performance. Finally, we present an approximation of the capacity when the PAPR is variable.

%%%%%%%%%%%%%%%%%%%%%%%%%%%%%%%%%%%%%%%%%%%%%%%%%%%%%%%%%%%%%%%%%%%%%%%%%%%%%%%%%%%%%%%%%%%%%%%%%%
%
%                                                                        @@@   @@
%          @@@                                                           @@@  @@@
%         @@@@@                                                          @@@
%         @@ @@@    @@@@@@@@@   @@@@@@@@@    @@@@@@@   @@ @@@@@     @@@@@@@@   @@  @@@   @@@
%        @@   @@    @@@@  @@@@  @@@@  @@@   @@@   @@@  @@@@  @@@   @@@  @@@@  @@@   @@@ @@@
%       @@@   @@@   @@@     @@  @@     @@@ @@@@@@@@@@  @@     @@  @@@    @@@   @@    @@@@
%       @@@@@@@@@   @@@     @@  @@     @@@ @@@@@@@@@@  @@     @@  @@@     @@   @@    @@@@
%      @@@      @@  @@@@   @@@  @@@    @@   @@         @@     @@   @@    @@@   @@   @@  @@
%     @@@       @@@ @@@@@@@@@   @@@@@@@@@    @@@@@@@@  @@     @@   @@@@@@@@@   @@  @@@   @@@
%     @@         @@ @@  @@@     @@  @@         @@@@     @     @@     @@@       @@  @      @@
%                   @@          @@
%                   @@          @@
%
%%%%%%%%%%%%%%%%%%%%%%%%%%%%%%%%%%%%%%%%%%%%%%%%%%%%%%%%%%%%%%%%%%%%%%%%%%%%%%%%%%%%%%%%%%%%%%%%%%
\appendix
\section*{Appendix A:  Proof of Theorem~\ref{T1}}
Eq. (\ref{E304}) can be explicitly expanded by
\begin{align} \label{E3082}
C=&\int_{\lambda}^{\frac{\lambda}{1-\lambda \ A \ \text{SNR}}}{\log{\left(\frac{t}{\lambda}\right)}  f_{{\left|{h}\right|^{2}}}\left(t\right)dt} \\ \nonumber
&+ \int_{\frac{\lambda}{1-\lambda \ A \ \text{SNR}}}^{\infty}{\log{\left(1+A \ \text{SNR} \ t\right)}f_{\left|{h}\right|^{2}}\left(t\right)dt }.
\end{align}
To simplify the notation, we define $\alpha(\lambda)=\frac{\lambda}{1-\lambda \ A \ \text{SNR}}$. Then, for the first term in the RHS of (\ref{E3082}),
\begin{align}\label{E308}
\int_{\lambda}^{\alpha(\lambda)} &\log{\left(\frac{t}{\lambda}\right)} f_{{\left|{h}\right|^{2}}}\left(t\right)dt   \nonumber \\
&\tilde{<}  \underset{t \in \left[ \lambda, \alpha(\lambda)  \right]}  {\max} f_{{\left|{h}\right|^{2}}}\left( t\right) \log{\left( \frac{\alpha(\lambda)}{\lambda} \right)} \left( \alpha(\lambda)- \lambda \right)  \\ \label{E319}
&\approx  f\left(\lambda\right) \lambda \ A \ \text{SNR} \lambda^2 A \ \text{SNR}     \\ \label{E320}
&= f\left(\lambda \right) \lambda^3 A^2 \text{SNR}^2 \\ \label{E321}
&\approx o\left(\text{SNR}\right).
\end{align}
Inequality (\ref{E308}) follows by substituting the integrand with its maximum value in the integral interval,  (\ref{E319}) by the properties of $\log{\left(1+x\right)} \approx x$, and $\frac{1}{1-x}-1 \approx x$, and (\ref{E320}) and (\ref{E321}) by Lemma~\ref{L2}.

The second term in the RHS of (\ref{E3082}) corresponds to
\begin{align}\label{E322}
\int_{\alpha(\lambda)}^{\infty}&{\log{\left(1+A \ \text{SNR} \ t\right)}f_{\left|{h}\right|^{2}} \left(t\right)dt } \nonumber \\
&\approx  \int_{\lambda}^{\infty}{\log{\left(1+A \ \text{SNR} \ t\right)}f_{\left|{h}\right|^{2}} \left(t\right)dt} \\ \label{E323}
&\approx A {\text{SNR}} \int_{\lambda}^{\infty}{t f_{\left|{h}\right|^{2}} \left(t\right)dt} \\ \label{E324}
&= A \ \text{SNR} \underset{\left|{h}\right|^{2} \ge \lambda }{\avr} {\left[ \left| {h}\right|^{2}\right]}.
\end{align}

Eq. (\ref{E322}) can be derived from the following proof:
\begin{align}
&\int_{\alpha(\lambda)}^{\infty}\log{\left(1+A \ \text{SNR} \ t\right)}f_{\left|{h}\right|^{2}} \left(t\right)dt \nonumber  \\
&\hspace{0.5cm}= \int_{\lambda}^{\infty}{\log{\left(1+A \ \text{SNR} \ t\right)}f\left(t\right)dt }  \\
&\hspace{1.5cm}-  \int_{\lambda}^{\alpha(\lambda)}{\log{\left(1+A \ \text{SNR} \ t\right)}f\left(t\right)dt }. \label{E325}
\end{align}

The first term in the RHS of Eq. (\ref{E325}) achieves the target in (\ref{E322}). Therefore, it is sufficient to prove the second term on the RHS of Eq. (\ref{E325}) is $o\left(\text{SNR}\right)$ provided (\ref{E323}) is satisfied. Following is the proof of this.
 \begin{align}\label{E326}
 &\int_{\lambda}^{\alpha(\lambda)}{\log{\left(1+A \ \text{SNR} \ t\right)}f\left(t\right)dt} \nonumber \\
 &\tilde{<}  \, \,  \underset{t \in \left[ \lambda, \alpha(\lambda)\right]} {\max} f\left( t\right) \left( \alpha(\lambda) -\lambda \right) \cdot  \log{\left(  1 + A \ \text{SNR} \cdot \alpha(\lambda) \right)} \\ \label{E327}
 &\approx f\left(\lambda \right)  \lambda^2 A \ \text{SNR} \lambda \ A \ \text{SNR} \\ \label{E328}
  &= o\left( \text{SNR}\right).
 \end{align}

Inequalities (\ref{E326}), (\ref{E327}) and (\ref{E328}) follow along similar lines as (\ref{E308}), (\ref{E319}) and (\ref{E320}).

Now it only remains to show that
\begin{equation}
 \int_{\lambda}^{\infty}{\log{\left(1+A \ \text{SNR} \  t\right)}f_{\left|{h}\right|^{2}}\left(t\right)dt } \approx A  \  \text{SNR} \underset{\left|{h}\right|^{2} \ge \lambda }{\avr} {\left[ \left| {h}\right|^{2}\right]}.
 \label{eq:inequal2}
\end{equation}

For any $ \epsilon > 0$, $t \in \left(\lambda, \infty\right)$, there exist some $\eta_t >0$, such that when $\text{SNR} \in \left[ 0, \eta_t \right]$, the following inequality holds:
\begin{equation}
\left | \frac {\log{\left( 1+t  A \ \text{SNR} \right)} }  {t A \ \text{SNR} } -1 \right| \leq \epsilon.
\label{eq:inequal1}
\end{equation}
This follows from the fact that $\log{\left(1+x \right)} \approx x $.

Define $\eta = \underset{t \in \left(\lambda, \infty\right)} {\min } \eta_t $. Then, when $\text{SNR} \in \left(0, \eta \right)$, from \eqref{eq:inequal1}, we have
\begin{align}
 &\int_{\lambda}^{\infty}{\left(1 - \epsilon\right)A \ \text{SNR} \ tf\left(t\right)dt} \nonumber\\
&\hspace{0.5cm}\leq \int_{\lambda}^{\infty}{\log{\left( 1+ A \ \text{SNR} \ t \right)  }f \left(t\right)  dt  } \\
&\hspace{0.5cm}\leq \int_{\lambda}^{\infty}{\left(1+ \epsilon \right) A \ \text{SNR} \ tf\left(t\right)dt}.
\end{align}
This is equivalently to show that
\begin{equation}
 \left| \frac{\int_{\lambda}^{\infty}{\log{\left( 1+ A \ \text{SNR} \ t \right)  }f \left(t\right)  dt  }}{\int_{\lambda}^{\infty}{t A \ \text{SNR}  f\left(t\right)dt}}  -1 \right | \leq  \epsilon.
\end{equation}
Since $\epsilon$ is arbitrarily chosen, then
\begin{align}
&\int_{\lambda}^{\infty}\hspace{-0.2cm}\log{\left(1+A \ \text{SNR} \ t\right)}f_{\left|{h}\right|^{2}}\left(t\right)dt  \approx  \int_{\lambda}^{\infty}\hspace{-0.2cm}{A \ \text{SNR} \ t}f_{\left|{h}\right|^{2}}\left(t\right)dt.
\end{align}
Therefore, the second term in the RHS of Eq.  (\ref{E3082}) dominates the first term.
From Lemma 2, we have $\lambda \approx F^{-1}\left(1-\frac{1}{A}\right)$. Hence, Eq.  \eqref{eq:inequal2} can be rewritten as
\begin{align}
 A \ \text{SNR} \underset{\left|{h}\right|^{2} \ge \lambda }{\avr} {\left[ \left| {h}\right|^{2}\right]}
&\approx\text{SNR} \ A  \int_{F^{-1}\left(1-\frac{1}{A}\right)}^{\infty}{{  t}f_{\left|{h}\right|^{2}}\left(t\right)dt } \\
&= \text{SNR} \  A \int_{1- \frac{1}{A}}^1  F^{-1}\left(t\right)dt \label{eq36 PAPR}
\end{align}
where \eqref{eq36 PAPR} is obtained after applying the Leibniz integral rule.
Theorem~\ref{T1} is proved.

\section*{Appendix B: Proof of Theorem~\ref{T2} }
From Lemma~(3) and Eq.  (\ref{E303}), the average power constraint at zero can be  expanded as:
\begin{IEEEeqnarray}{c}\label{E333}
\text{SNR} \approx e^{-\lambda}\left( \frac{1}{\lambda ^2} -\left( \frac{1}{\lambda } - \mathcal{A}\left(\text{SNR}\right) \text{SNR} \right)^2 e^{-l} + o\left( \frac{1}{\lambda^2}  \right) \right)
\nonumber \\
\end{IEEEeqnarray}

Similarly, the capacity can be expanded as:
\begin{equation}\label{E334}
 C \approx e^{-\lambda}\left( \frac{1}{\lambda}  -  \left(   \frac{1}{\lambda} - \mathcal{A}\left(\text{SNR}\right)\text{SNR} - \frac{l}{\lambda^2}  \right) e^{-l}  \right).
\end{equation}

1) If $l_{0}  >  0$, Eq. \eqref{E333} and \eqref{E334} can be asymptotically estimated as
\begin{equation}\label{E335}
\text{SNR} \approx \frac{e^{-\lambda}}{\lambda^2}  \left( 1- e^{-l_{0}}  \right),
\end{equation}
\begin{equation}\label{E336}
 C \approx \frac{e^{-\lambda}}{\lambda} \left(1- e^{-l_{0}} \right).
\end{equation}

Therefore, from the results in~\cite{Rezki2012}, we can see that $\lambda$ in Eq.  (\ref{E335}) is actually
\begin{equation}
 \lambda \approx \log {\frac{1}{\text{SNR}}.}
 \label{eq:lambda}
\end{equation}

Substituting \eqref{eq:lambda} into \eqref{E336}, we get
\begin{equation}
C \approx  \text{SNR} \log{\frac{1}{\text{SNR}} }.
\end{equation}

2) If $l_{0} = 0$, Eq. \eqref{E333} and \eqref{E334} can then be asymptotically estimated as
\begin{equation}\label{E337}
\text{SNR} \approx \mathcal{A}\left(\text{SNR} \right)\text{SNR} e^{-\lambda},
\end{equation}
\begin{equation}\label{E338}
 C \approx \lambda \ \mathcal{A}\left(\text{SNR}\right)\text{SNR} e^{-\lambda}.
\end{equation}

Equalilties \eqref{E337} and \eqref{E338} follow by the series expansion ${e^{-l}=1-l+\frac{l^{2}}{2}+o(l^2)}$, and by choosing the dominating terms. Hence, the capacity can be expressed as:
\begin{equation}
C \approx \text{SNR}\log{ \mathcal{A}\left(\text{SNR}\right)}.
\end{equation}
Theorem 2 is proved. 
\bibliographystyle{IEEEtran}
\bibliography{references}

\begin{thebibliography}{10}
\providecommand{\url}[1]{#1}
\csname url@samestyle\endcsname
\providecommand{\newblock}{\relax}
\providecommand{\bibinfo}[2]{#2}
\providecommand{\BIBentrySTDinterwordspacing}{\spaceskip=0pt\relax}
\providecommand{\BIBentryALTinterwordstretchfactor}{4}
\providecommand{\BIBentryALTinterwordspacing}{\spaceskip=\fontdimen2\font plus
\BIBentryALTinterwordstretchfactor\fontdimen3\font minus
  \fontdimen4\font\relax}
\providecommand{\BIBforeignlanguage}[2]{{%
\expandafter\ifx\csname l@#1\endcsname\relax
\typeout{** WARNING: IEEEtran.bst: No hyphenation pattern has been}%
\typeout{** loaded for the language `#1'. Using the pattern for}%
\typeout{** the default language instead.}%
\else
\language=\csname l@#1\endcsname
\fi
#2}}
\providecommand{\BIBdecl}{\relax}
\BIBdecl

\bibitem{Verdu2002}
S.~Verdu, ``Spectral efficiency in the wideband regime,'' \emph{IEEE
  Transactions on Information Theory}, vol.~48, no.~6, pp. 1319--1343, June
  2002.

\bibitem{Sboui2013b}
L.~Sboui, Z.~Rezki, and M.-S. Alouini, ``Capacity of spectrum sharing cognitive
  radio systems over {N}akagami fading channels at low {SNR},'' in \emph{Proc.
  of the IEEE International Conf. on Comm. (ICC'13), Budapest, Hungary}, June
  2013, pp. 5674--5678.

\bibitem{Sboui2014b}
L.~Sboui, Z.~Rezki, and M.~S. Alouini, ``Achievable rate of spectrum sharing
  cognitive radio systems over fading channels at low-power regime,''
  \emph{IEEE Transactions on Wireless Communications}, vol.~13, no.~11, pp.
  6461--6473, Nov. 2014.

\bibitem{Yao2005}
Y.~Yao, X.~Cai, and G.~B. Giannakis, ``On energy efficiency and optimum
  resource allocation of relay transmissions in the low-power regime,''
  \emph{IEEE Transactions on Wireless Communications}, vol.~4, no.~6, pp.
  2917--2927, Nov. 2005.

\bibitem{Sboui2017}
L.~Sboui, Z.~Rezki, and M.~S. Alouini, ``Achievable rates of cognitive radio
  networks using multilayer coding with limited {CSI},'' \emph{IEEE
  Transactions on Vehicular Technology}, vol.~66, no.~1, pp. 395--405, Jan.
  2017.

\bibitem{shannon48_1}
C.~E. Shannon, ``A mathematical theory of communication,'' vol.~27, pp.
  379--423 and 623--656, Jul. 1948.

\bibitem{kennedy_69}
R.~S. Kennedy, \emph{Fading dispersive communication channels}.\hskip 1em plus
  0.5em minus 0.4em\relax Wiley-Interscience, 1969.

\bibitem{telatar_00}
I.~E. Telatar and D.~N.~C. Tse, ``Capacity and mutual information of wideband
  multipath fading channels,'' \emph{IEEE Transactions on Information Theory},
  vol.~46, no.~4, pp. 1384--1400, 2000.

\bibitem{medard_02}
M.~M{\'e}dard and R.~G. Gallager, ``Bandwidth scaling for fading multipath
  channels,'' \emph{IEEE Transactions on Information Theory}, vol.~48, no.~4,
  pp. 840--852, 2002.

\bibitem{verdu_02}
S.~Verd{\'u}, ``Spectral efficiency in the wideband regime,'' \emph{IEEE
  Transactions on Information Theory}, vol.~48, no.~6, pp. 1319--1343, 2002.

\bibitem{Borade2010}
S.~Borade and L.~Zheng, ``Wideband fading channels with feedback,'' \emph{IEEE
  Transactions on Information Theory}, vol.~56, no.~12, pp. 6058--6065, Dec.
  2010.

\bibitem{Rezki2012}
Z.~Rezki and M.-S. Alouini, ``On the capacity of {N}akagami-m fading channels
  with full channel state information at low {SNR},'' \emph{IEEE Wireless
  Communication Letters}, vol.~1, no.~3, pp. 253--256, June 2012.

\bibitem{porrat_07}
D.~Porrat, ``Information theory of wideband communications.'' \emph{IEEE
  Communications Surveys and Tutorials}, vol.~9, no. 1-4, pp. 2--16, 2007.

\bibitem{Tarokh2000}
V.~Tarokh and H.~Jafarkhani, ``On the computation and reduction of the
  peak-to-average power ratio in multicarrier comm.'' \emph{IEEE Trans. on
  Comm.}, vol.~48, no.~1, pp. 37--44, Jan. 2000.

\bibitem{smith_71}
J.~G. Smith, ``The information capacity of amplitude-and variance-constrained
  sclar {G}aussian channels,'' \emph{Information and Control}, vol.~18, no.~3,
  pp. 203--219, 1971.

\bibitem{shamai_95}
S.~Shamai and I.~Bar-David, ``The capacity of average and peak-power-limited
  quadrature {G}aussian channels,'' \emph{IEEE Transactions on Information
  Theory}, vol.~41, no.~4, pp. 1060--1071, 1995.

\bibitem{abou_01}
I.~C. Abou-Faycal, M.~D. Trott, and S.~Shamai, ``The capacity of discrete-time
  memoryless {R}ayleigh-fading channels,'' \emph{IEEE Transactions on
  Information Theory}, vol.~47, no.~4, pp. 1290--1301, 2001.

\bibitem{gursoy05_1}
M.~C. Gursoy, H.~V. Poor, and S.~Verd{\'u}, ``Noncoherent {R}ician fading
  channel-part ii: spectral efficiency in the low-power regime,'' \emph{IEEE
  Transactions on Wireless Communications}, vol.~4, no.~5, pp. 2207--2221,
  2005.

\bibitem{gursoy05_2}
------, ``The noncoherent {R}ician fading channel-part i: structure of the
  capacity-achieving input,'' \emph{IEEE Transactions on Wireless
  Communications}, vol.~4, no.~5, pp. 2193--2206, 2005.

\bibitem{dytso_17}
A.~Dytso, M.~Goldenbaum, S.~Shamai, and H.~V. Poor, ``Upper and lower bounds on
  the capacity of amplitude-constrained {MIMO} channels,'' in \emph{Proc. of
  the IEEE Global Commun. Conf. (GLOBECOM'17)}, 2017, pp. 1--6.

\bibitem{lapidoth_09}
A.~Lapidoth, S.~M. Moser, and M.~A. Wigger, ``On the capacity of free-space
  optical intensity channels,'' \emph{IEEE Transactions on Information Theory},
  vol.~55, no.~10, pp. 4449--4461, 2009.

\bibitem{moser_18}
S.~M. Moser, L.~Wang, and M.~Wigger, ``Capacity results on multiple-input
  single-output wireless optical channels,'' \emph{IEEE Transactions on
  Information Theory}, 2018.

\bibitem{chaaban18_1}
A.~Chaaban, Z.~Rezki, and M.-S. Alouini, ``Capacity bounds and high-{SNR}
  capacity of {MIMO} intensity-modulation optical channels,'' \emph{IEEE
  Transactions on Wireless Communications}, vol.~17, no.~5, pp. 3003--3017,
  2018.

\bibitem{chaaban18_2}
------, ``Low-{SNR} asymptotic capacity of {MIMO} optical intensity channels
  with peak and average constraints,'' \emph{IEEE Transactions on
  Communications}, 2018.

\bibitem{subramanian_02}
V.~G. Subramanian and B.~Hajek, ``Broad-band fading channels: {S}ignal
  burstiness and capacity,'' \emph{IEEE Transactions on Information Theory},
  vol.~48, no.~4, pp. 809--827, 2002.

\bibitem{Goldsmith1997}
A.~Goldsmith and P.~Varaiya, ``Capacity of fading channels with channel side
  information,'' \emph{IEEE Transactions on Information Theory}, vol.~43,
  no.~6, pp. 1986--1992, Nov. 1997.

\bibitem{Khojastepour2004}
M.~Khojastepour and B.~Aazhang, ``The capacity of average and peak power
  constrained fading channels with channel side information,'' in \emph{Proc.
  of the IEEE Wireless Comm. and Networking Conf. (WCNC'04)}, vol.~1, Mar.
  2004, pp. 77--82.

\end{thebibliography}

\begin{IEEEbiography}[{\includegraphics[width=1in, height=1.25in, clip,keepaspectratio]{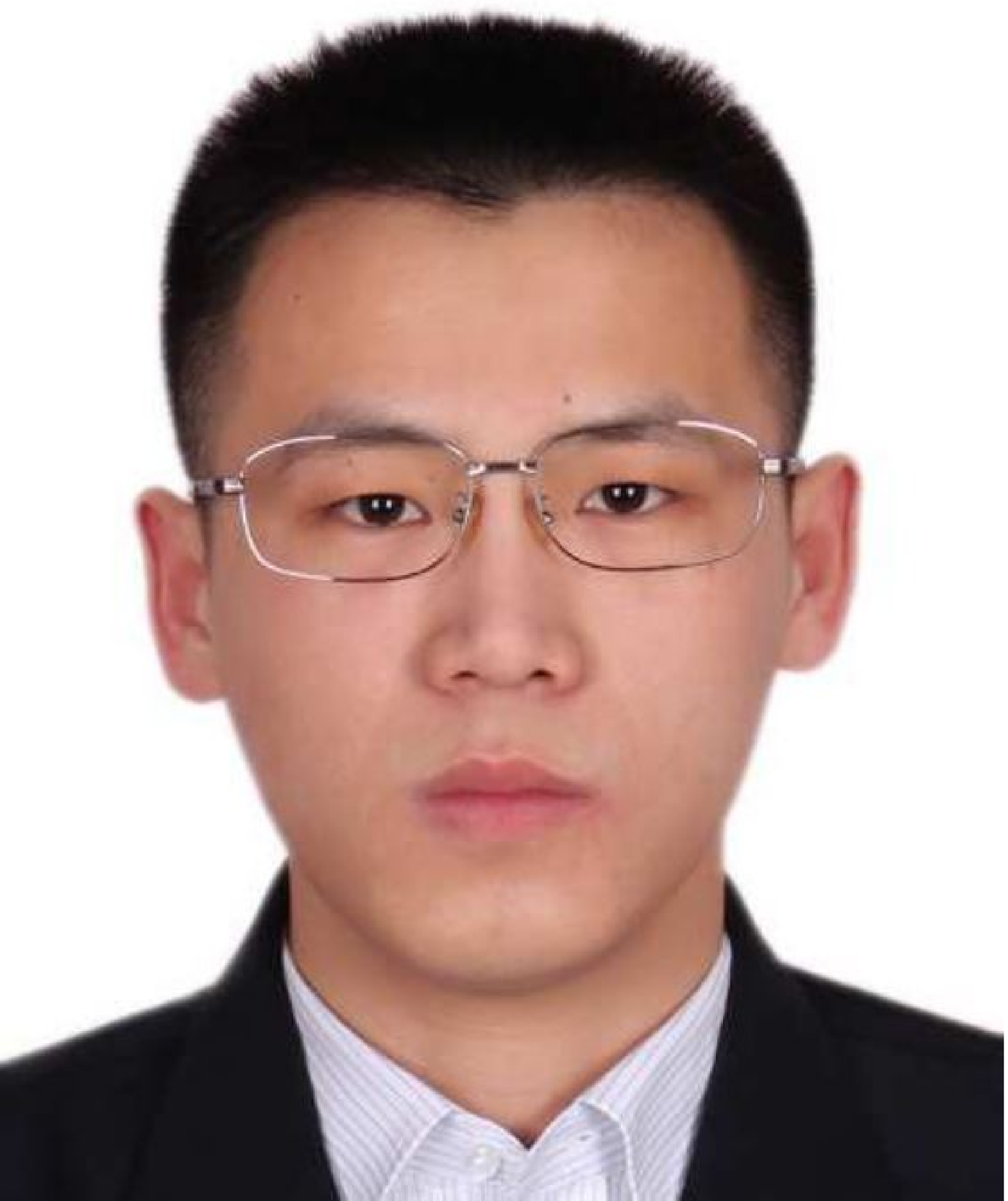}}]{Longguang Li}
received the M.Sc degree in Electrical Engineering in 2016 from Shanghai Jiao Tong University, Shanghai, China.
He is currently a Ph.D. candidate at Telecom ParisTech. His research interests are in information theory and wireless communications.
\end{IEEEbiography}

\begin{IEEEbiography}[{\includegraphics[width=1in,height=1.25in,clip,keepaspectratio]{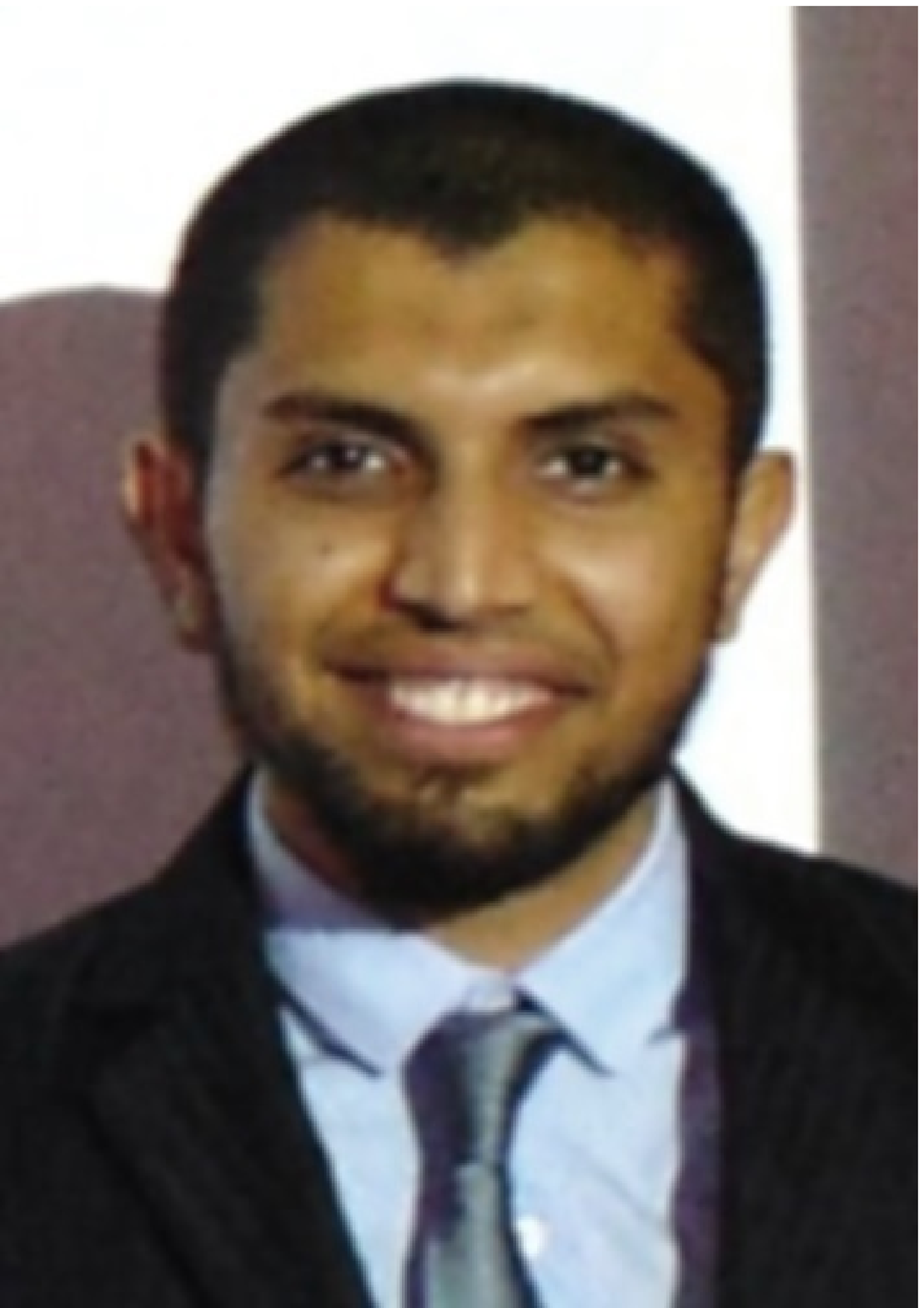}}]{Lokman Sboui}(S'11-M'17) was born in Cairo, Egypt. He received the Diplome d'Ing\'enieur degree with honors from Ecole Polytechnique de Tunisie (EPT), La Marsa, Tunisia, in 2011, the M.Sc. degree from King Abdullah University of Science and Technology (KAUST) in 2013, and the Ph.D. degree from King Abdullah University of Science and Technology (KAUST) in 2017. He is currently a Postdoctoral Researcher in the Department of Electrical Engineering at  \'Ecole de Technologie Sup\'erieure (\'ETS), Montréal, Canada. His current research interests include: performance of cognitive radio systems, low SNR communication, energy efficient power allocation.
\end{IEEEbiography}

\begin{IEEEbiography}[{\includegraphics[width=1in,height=1.25in,clip,keepaspectratio]{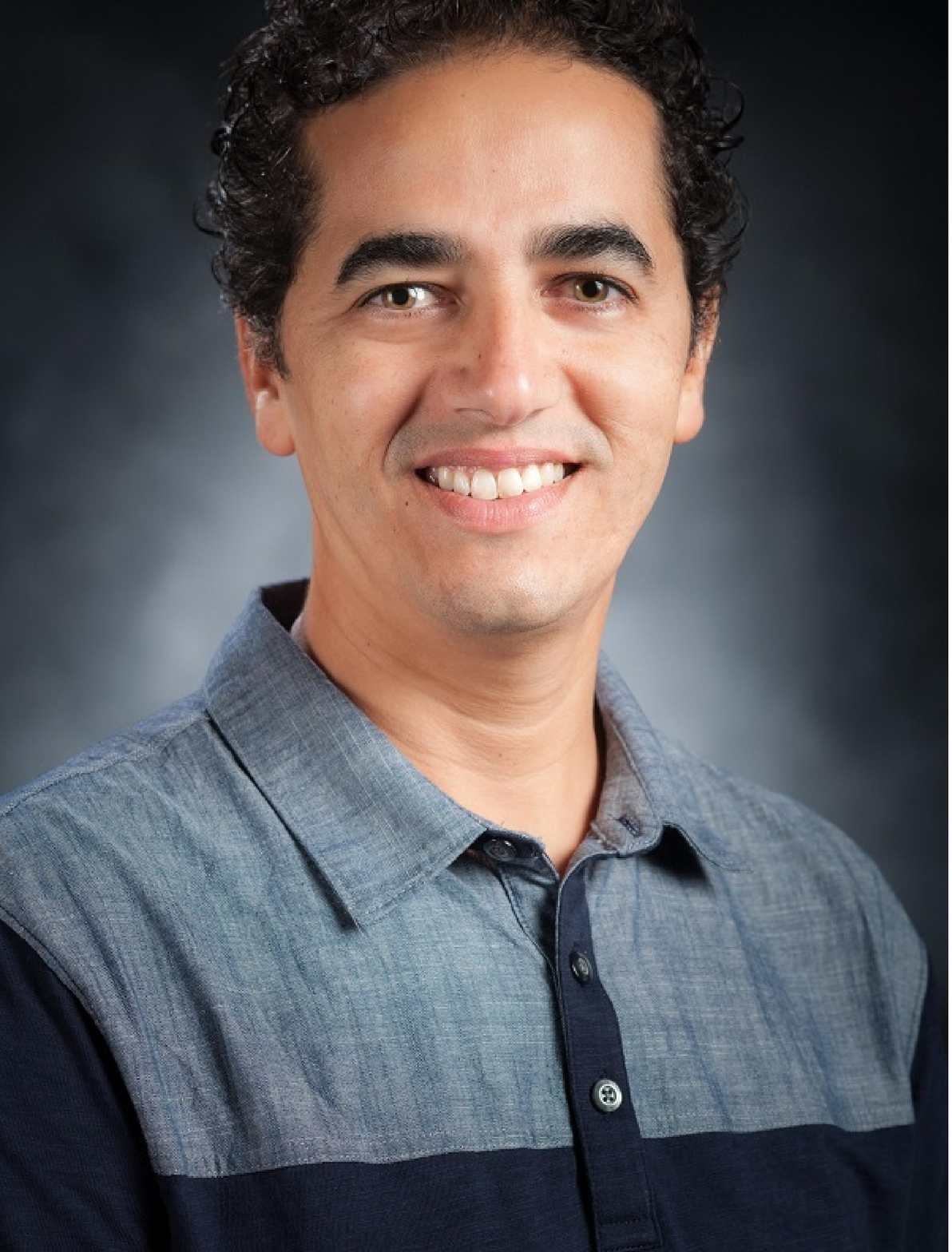}}] {Zouheir Rezki} (S'01-M'08-SM'13) was born in Casablanca, Morocco. He received the Dipl\^ome d'Ing\'enieur degree from the \'Ecole Nationale de l'Industrie Min\'erale (ENIM), Rabat, Morocco, in 1994, the M.Eng. degree from  \'Ecole de Technologie Sup\'erieure, Montreal, Qu\'ebec, Canada, in 2003, and the Ph.D. degree in electrical engineering from  \'Ecole Polytechnique, Montreal, Qu\'ebec, Canada, in 2008. After a few years of experience as a postdoctoral fellow and a research scientist at KAUST, he joined University of Idaho as an Assistant Professor in the ECE Department.
\end{IEEEbiography}

\begin{IEEEbiography}[{\includegraphics[width=1in,height=1.25in,keepaspectratio]{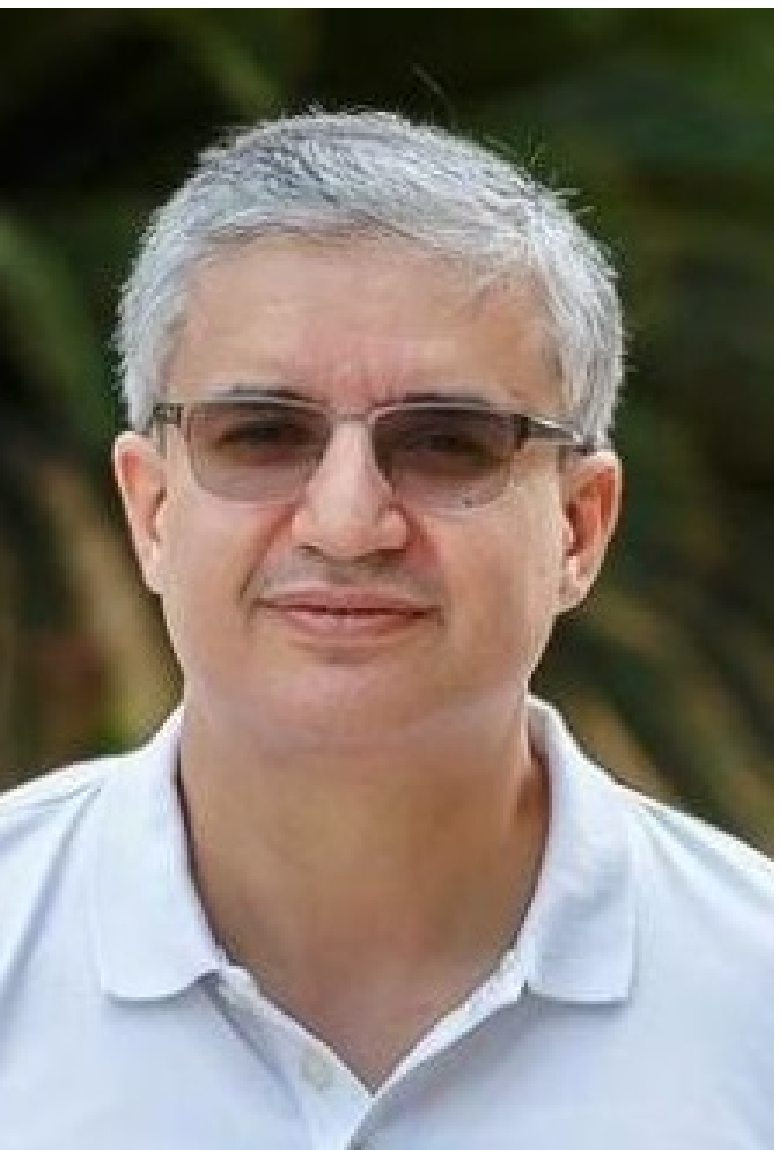}}]{Mohamed-Slim Alouini}
(S'94-M'98-SM'03-F'09)  was born in Tunis, Tunisia. He received the Ph.D. degree in Electrical Engineering
from the California Institute of Technology (Caltech), Pasadena,
CA, USA, in 1998. He served as a faculty member in the University of Minnesota,
Minneapolis, MN, USA, then in the Texas A$\&$M University at Qatar,
Education City, Doha, Qatar before joining King Abdullah University of
Science and Technology (KAUST), Thuwal, Makkah Province, Saudi
Arabia as a Professor of Electrical Engineering in 2009. His current
research interests include the modeling, design, and
performance analysis of wireless communication systems.
\end{IEEEbiography}

\end{document}